\documentclass[10pt,aps,reprint,prb,floatfix,superscriptaddress,longbibliography]{revtex4-2}
\pdfoutput=1
\usepackage{graphicx,latexsym}
\usepackage{dcolumn,amsfonts}
\usepackage{amssymb,amsmath,bm}

\usepackage[breaklinks=true,pdfencoding=auto]{hyperref}

\usepackage{natbib}
\usepackage{balance}

\hypersetup{
	colorlinks   = true, 
	urlcolor     = blue, 
	linkcolor    = blue, 
	citecolor    = red 
}

\usepackage{color}
\usepackage{ulem}

\begin{document}

	\title{Spin-phase transition in an array of quantum rings controlled by cavity photons}

\author{Vidar Gudmundsson}
	\email{vidar@hi.is}
	\affiliation{Science Institute, University of Iceland, Dunhaga 3, IS-107 Reykjavik, Iceland}
	\author{Vram Mughnetsyan}
	\email{vram@ysu.am}
	\affiliation{Department of Solid State Physics, Yerevan State University, Alex Manoogian 1, 0025 Yerevan, Armenia}
	\author{Hsi-Sheng Goan}
	\email{goan@phys.ntu.edu.tw}
    \affiliation{Department of Physics and Center for Theoretical Physics, National Taiwan University, Taipei 106319, Taiwan}
	\affiliation{Center for Quantum Science and Engineering, National Taiwan University, Taipei 106319, Taiwan}
	\affiliation{Physics Division, National Center for Theoretical Sciences, Taipei 106319, Taiwan}
	\author{Jeng-Da Chai}
	\email{jdchai@phys.ntu.edu.tw}
	\affiliation{Department of Physics and Center for Theoretical Physics, National Taiwan University, Taipei 106319, Taiwan}
	\affiliation{Center for Quantum Science and Engineering, National Taiwan University, Taipei 106319, Taiwan}
	\affiliation{Physics Division, National Center for Theoretical Sciences, Taipei 106319, Taiwan}
	\author{Nzar Rauf Abdullah}
	\email{nzar.r.abdullah@gmail.com}
	\affiliation{Physics Department, College of Science,
		University of Sulaimani, Kurdistan Region, Iraq}
	\author{Chi-Shung Tang}
	\email{cstang@nuu.edu.tw}
	\affiliation{Department of Mechanical Engineering, National United University, Miaoli 360302, Taiwan}
	\author{Valeriu Moldoveanu}
	\email{valim@infim.ro}
	\affiliation{National Institute of Materials Physics, PO Box MG-7, Bucharest-Magurele,
		Romania}
	\author{Andrei Manolescu}
	\email{manoles@ru.is}
	\affiliation{Department of Engineering, Reykjavik University, Menntavegur
		1, IS-102 Reykjavik, Iceland}

%

\begin{abstract}
We model a spin-phase transition in a two-dimensional square array, or a lateral superlattice,
of quantum rings in an external perpendicular homogeneous magnetic field. The electron system is
placed in a circular cylindrical far-infrared photon cavity with a single circularly symmetric
photon mode. Our numerical results reveal that the spin ordering of the two-dimensional electron
gas in each quantum ring can be influenced or controlled by the electron-photon coupling strength
and the energy of the photons. The Coulomb interaction between the electrons is described by a
spin-density functional approach, but the para- and the diamagnetic electron-photon interactions
are modeled via a configuration interaction formalism in a truncated many-body Fock-space,
which is updated in each iteration step of the density functional approach.
In the absence of external electromagnetic pulses this spin-phase transition is replicated
in the orbital magnetization of the rings. The spin-phase transition can be suppressed by
a strong electron-photon interaction. 
{In addition, fluctuations in the spin configuration are
found in dynamical calculations, where the system is excited by a time-dependent
coupling scheme to a cylindrical cavity mode
for emphasizing the diamagnetic electron-photon interaction 
not leading to simple electrical 
dipole oscillations}.
\end{abstract}

\maketitle
%
%

\section{Introduction}
Researchers have been exploring and
modeling \cite{Flick2017,FlickRiveraNarang+2018+1479+1501,10.1063/1.5142502,Buchholz2019}
{the effects of photon cavity fields on chemical systems and reactions}
\cite{PhysRevLett.116.238301,Du2019,https://doi.org/10.1002/anie.201107033},
molecules \cite{Schafer2021ShiningLO}, and
nanoscale electron systems \cite{PhysRevB.101.075301,PhysRevB.101.205140}
in order to modify or control their properties.
On the theoretical side, several methods have been used, 
{some of which have been} gathered under the
umbrella concept of quantum electrodynamical density functional theory
(QEDFT) \cite{doi:10.1021/acsphotonics.7b01279,flick2021simple,10.1063/5.0123909,PhysRevLett.133.096401}.
Concurrently, this interest has been extended to large systems with two-dimensional electron gas
(2DEG) in heterostructures. Experiments have indicated that the 2DEG in GaAs heterostructures
is especially advantageous for fundamental research on the 2DEG-cavity photon interactions due
to its high purity and polarizability \cite{Zhang1005:2016,doi:10.1126/science.1216022}.
The free homogeneous 2DEG placed in a parallel plate cavity has been modeled
\cite{PhysRevResearch.4.013012,PhysRevLett.123.047202,PhysRevB.105.205424}, as well as
the 2DEG modulated in a superlattice and subjected to the static electronic Coulomb interaction, both
in the static \cite{PhysRevB.106.115308,mughnetsyan2023magnetic,PhysRevB.109.235306},
or in the dynamic case after short excitation
pulse \cite{PhysRevB.108.115306,PhysRevB.110.205301}.

Previously, the present group of researchers has modeled the modulated 2DEG in a
square array of quantum dots interacting with the quantized TE$_{011}$ mode of a circular cylindrical
cavity. That was done in order to strengthen the diamagnetic electron-photon interaction of the system
placed in a homogeneous external magnetic field promoting persistent circular or
rotational currents. The model calculations were accomplished for the static system
using the quantum electrodynamic density functional theory tensor product (QED-DFT-TP)
formalism presented by Malave
{\it et al.}\ \cite{10.1063/5.0123909} adapted to the 2DEG and the external magnetic
field \cite{PhysRevB.109.235306}. Particularly, the excitations of the system with a short
pulse-formed modulation of the electron-photon interaction in a self-consistent real-time approach
showed how diamagnetic two-photon transitions are strongly promoted in the
system \cite{PhysRevB.110.205301}.

This has directed us to consider a system with the 2DEG in a lateral superlattice of quantum rings.
{
The term superlattice is used to emphasize that this periodic man-made structure has no relation
to the underlying structure of the atomic crystal lattice}.
There are two reasons for the present study: i) the ring array may be more susceptible to the symmetry
of the TE$_{011}$ mode of a cylindrical cavity, and ii) the exchange correlations between the rings
may play a special role in the system.
This last point is supported by our earlier findings for the dot system, i.e.\ an array
of quantum dots, where the results of the calculations using the QED-DFT-TP \cite{PhysRevB.109.235306},
or the QEDFT \cite{PhysRevB.106.115308} clearly show that,
an increased electron-photon interaction weakens the Coulomb exchange interactions considerably.
Last but not least, we believe that a quantum ring array is a suitable platform to further test
the QED-DFT-TP method.

{In the present investigation we find that the Coulomb exchange forces in a ring array are stronger
than in a corresponding array of quantum dots \cite{PhysRevB.109.235306,PhysRevB.110.205301}. 
For a ring array with 2 electrons in each ring the electron system of each ring stays in a
triplet spin state, even at low magnetic field. This is neither the case for individual quantum
dots \cite{Pfannkuche93:2244} nor arrays thereof with the same lattice length \cite{PhysRevB.109.235306}.
For a ring array with 3 electrons in each unit, we find a fully spin polarized state for low electron-photon
coupling, but we observe a spin-phase transition as the electron-photon coupling is increased. We map out
how this varies with the energy of the cavity photons. 
Dynamic excitation of the electrons in a ring array promoting diamagnetic transitions
does not change its spin structure but reveals temporal spin-fluctuations close to the regime 
of the spin-phase transition.}

The paper is organized as follows: In Sec.\ \ref{Model} we briefly describe the model,
which was already introduced in our previous works.
The results and discussion thereof are found in Sec.\ \ref{Results}, with the conclusions
drawn in Sec.\ \ref{Conclusions}. Appendix \ref{Tech-details} contains technical details
of the methodology used for the modeling.

\section{Model}
\label{Model}
The mutual Coulomb interaction of the electrons in the 2DEG are described within
spin density functional theory using an exact direct term, and approximate exchange
correlations terms detailed in the Appendix of Ref.\ \cite{PhysRevB.106.115308}.
The extension to systems in homogeneous external magnetic field was presented by
Lubin {\it et al.}\ and Koskinen {\it et al.}\ \cite{Lubin97:10373,Koskinen97:1389}
and built on development by
Barth and Hedin \cite{Barth_1972}, and Tanatar and Ceperley \cite{Tanatar89:5005}.
The 2DEG in the $xy$-plane is in an external homogeneous magnetic field $\bm{B} = B{\bm{e}_z}$
described by the vector potential $\bm{A} = (B/2)(-y,x)$. The external magnetic field
introduces the natural length scale, the magnetic length $l = (\hbar c/(eB))^{1/2}$, and
an energy scale, the cyclotron energy $\hbar\omega_c = (eB/(m^*c))$.

\begin{figure}[htb]
    \includegraphics[width=0.48\textwidth]{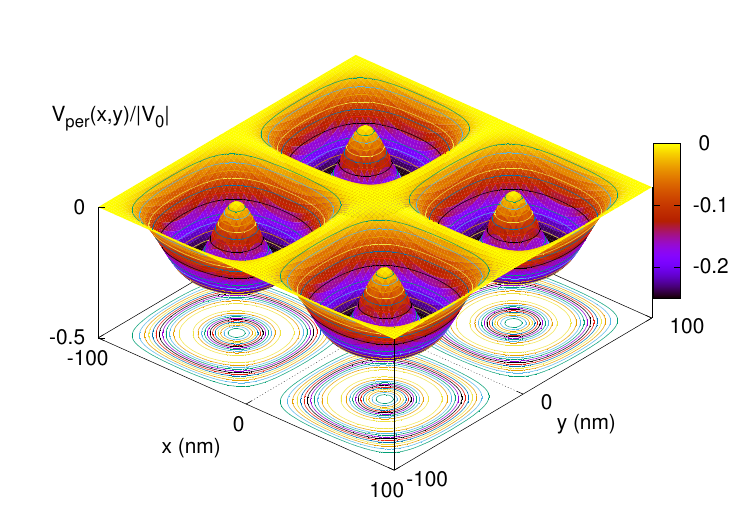}
    \caption{Four unit cells of the potentials defining the two-dimensional array of quantum rings.
        The superlattice length $L = 100$ nm. {The color scale for $V_\mathrm{per}(x,y)/|V_0|$ is
        shown on the right side. Equipotential contour lines of the potential are shown at both the surface and
        the base. The $z$-range [-0.5:0.0] is covered by 16 equipotentials with same color on the
        surface and the base. The $x$ and the $y$ axes are indicated on the base with dashed thin
        black lines}.}
    \label{Vxy}
\end{figure}

The potential defining the square array, or lateral superlattice, of rings is
\begin{align}
       V_\mathrm{per}(\bm{r}) = &-V_0\left[\sin \left(\frac{g_1x}{2} \right)
       \sin\left(\frac{g_2y}{2}\right) \right]^2 \nonumber\\
                                &+V_0\left[\sin \left(\frac{g_1x}{2} \right)
       \sin\left(\frac{g_2y}{2}\right) \right]^4,
\label{Vper}
\end{align}
with $V_0 = 64.0$ meV. The potential is displayed in Fig.\ \ref{Vxy} for
4 unit cells of the extended superlattice hosting the 2DEG.
The square superlattice of quantum rings is spanned by the spatial vectors
$\bm{R}=n\bm{l}_1+m\bm{l}_2$ with $n,m\in \bm{Z}$,
where the unit vectors are $\bm{l}_1 = L\bm{e}_x$ and $\bm{l}_2 = L\bm{e}_y$.
The reciprocal lattice is spanned by $\bm{G} = G_1\bm{g}_1 + G_2\bm{g}_2$ with
$G_1, G_2\in \mathbf{Z}$ and its unit vectors $\bm{g}_1 = 2\pi\bm{e}_x/L$
and $\bm{g}_2 = 2\pi\bm{e}_y/L$. The period of the superlattice is $L = 100$ nm,
setting another length scale competing with the magnetic length $l$
\cite{Hofstadter76:2239}. {Regular arrays or lateral superlattices of
quantum dots and rings have been made in the past with several methods
\cite{Heitmann93:56,WangHanHopkinson+2023+1469+1479}.}

The 2DEG is placed in the center of a circular cylindrical {far-infrared (FIR)} photon cavity.
The long wavelength approximation for the vector potential for the TE$_{011}$ cavity
photon mode was derived in the Appendix of Ref.\ \cite{PhysRevB.109.235306} to be
\begin{equation}
    {\bm A}_\gamma (\bm{r}) = \bm{e}_\phi {\cal A}_\gamma\left(a^\dagger_\gamma + a_\gamma  \right)
    \left(\frac{r}{l} \right),
\label{Agamma}
\end{equation}
where $\bm{e}_\phi$ is the unit angular vector for the polar coordinates, $(r,\phi)$.
The vector potential (\ref{Agamma}) exhibits the same spatial form as the vector potential
${\bm A}$ of the external homogeneous magnetic field $\bm{ B} = B\bm{e}_z$ \cite{PhysRevB.109.235306}.
$a^\dagger_\gamma$ and $a_\gamma$ are the photon creation and annihilation operators of the
FIR cavity photon mode, respectively. {The label $\gamma$ is used to note variables and operators
of the photon field}.

The electrons of the 2DEG interact with the cavity photon mode according to
\begin{align}
     H_\mathrm{int} = \frac{1}{c}\int_{\mathbf{R}^2} d\bm{r}\; &
     {\bm J}({\bm r})\cdot{\bm A}_\gamma (\bm{r}) \nonumber\\
     +& \frac{e^2}{2m^*c}\int_{\mathbf{R}^2} d\bm{r}\;
     n_\mathrm{e}(\bm{r})A^2_\gamma(\bm{r}),
     \label{e-g}
\end{align}
where both the para- and the diamagnetic parts of the electron-photon interaction are
accounted for. In the static part of the calculations the electron and the current
densities are defined as
\begin{align}
    n_\mathrm{e}(\bm{r})
    = \frac{1}{(2\pi)^2}\sum_{\bm{\alpha}\sigma}
    \int^{\pi}_{-\pi}d\bm{\theta}\; \left|\psi_{\bm{\alpha\theta}\sigma}(\bm{r}) \right|^2
    f(E_{\bm{\alpha\theta}\sigma}-\mu),
    \label{ne}
\end{align}
\begin{align}
    \bm{J}_{i}(\bm{r}) = \frac{-e}{m^*(2\pi)^2}\sum_{{\bm \alpha}\sigma}\int_{-\pi}^{\pi} d\bm{\theta}\;
    \Re&\left\{ \psi_{\bm{\alpha\theta}\sigma}^*(\bm{r})\bm{\pi}_i \psi_{\bm{\alpha\theta}\sigma}(\bm{r}) \right\}\nonumber\\
    &f(E_{\bm{\alpha\theta}\sigma}-\mu),
    \label{currD}
\end{align}
respectively, for $i=x$ or $y$.
$f$ notes the equilibrium Fermi distribution, $\mu$ is the chemical potential of the
electron system, and
\begin{equation}
    \bm{\pi} = \left(\bm{p}+\frac{e}{c}\bm{A} \right).
\label{pi}
\end{equation}
The wavefunctions $\psi_{\bm{\alpha\theta}\sigma}$ correspond to the states
$\{|\bm{\alpha\theta}\sigma)\}$ of the self-consistently diagonalized static Hamiltonian
of the electron-photon system with the energy spectrum $E_{\bm{\alpha\theta}\sigma}$.

In the Appendix of Ref.\ \cite{PhysRevB.109.235306}
the electron-photon interaction (\ref{e-g}) is rewritten in terms of the photon creation
and annihilation operators as
\begin{align}
    \label{e-gIxIyN}
    H_\mathrm{int} &= g_\gamma \hbar\omega_c \left\{ lI_x + lI_y\right\} \left(a^\dagger_\gamma + a_\gamma\right)\\
    &+ g^2_\gamma \hbar\omega_c {\cal N}\left\{\left(a^\dagger_\gamma a_\gamma + \frac{1}{2}\right)
    +\frac{1}{2}\left(a^\dagger_\gamma a^\dagger_\gamma + a_\gamma a_\gamma\right)\right\}\nonumber,
\end{align}
where conveniently and importantly, the constants $I_x$, $I_y$, and ${\cal N}$ turn out as
functionals the electron charge- and current densities. Previous calculations with the
model for a square array of quantum dots interacting with the TE$_{011}$ cylindrical mode show
the importance of retaining all the terms of the diamagnetic part of the interaction,
the second line in Eq.\ (\ref{e-gIxIyN}) \cite{PhysRevB.110.205301}.
The dimensionless electron-photon coupling constant emerging in the derivation is
\begin{equation}
    g_\gamma = \left\{ \left( \frac{e{\cal A}_\gamma}{c} \right) \frac{l}{\hbar} \right\}.
\end{equation}
The energy of the photons in the single cavity mode is $E_\gamma = \hbar\omega_\gamma$.

In order to explore the dynamical evolution of the system, it is excited with
a short pulse-like modulation of the electron-photon interactions.
\begin{align}
    \label{Ht}
    H_\mathrm{ext}(t) &= F(t)
    \biggl[ g_\gamma \hbar\omega_c\left\{lI_x + lI_y \right\} \left( a^\dagger_\gamma + a_\gamma \right)\\
    + &\left. g_\gamma^2 \hbar\omega_c {\cal N} \left\{ \left( a^\dagger_\gamma a_\gamma + \frac{1}{2}\right)
    + \frac{1}{2}\left( a^\dagger_\gamma a^\dagger_\gamma + a_\gamma a_\gamma  \right) \right\} \right]\nonumber
\end{align}
with
\begin{equation}
    F(t) = \left( \frac{V_t}{\hbar\omega_c}\right) (\Gamma t)^2 \exp{(-\Gamma t)}\cos{(\omega_\mathrm{ext} t)}.
    \label{ft}
\end{equation}
The time-evolution of the system is calculated self-consistently with the Liouville-von Neumann
equation for the density, or the probability, operator $\rho^{\bm{\theta}} (t)$ for each
$\bm{\theta}$-point in the first Brillouin zone of the reciprocal lattice.
\begin{equation}
    i\hbar\partial_t \rho^{\bm{\theta}} (t) = \left[ H[\rho^{\bm{\theta}}(t)], \rho^{\bm{\theta}}(t)\right].
    \label{L-vN}
\end{equation}
That is possible as, neither any term in the electron-photon interaction (\ref{e-gIxIyN}), nor the
excitation (\ref{Ht}) couple different $\bm{\theta}$ points in the reciprocal lattice.
Self-consistency refers to that the direct Coulomb and the exchange-correlation functionals, depending
on the electron density, have to be updated in each iteration within each time step together with the
functionals $I_x$, $I_y$, and ${\cal N}$ of the electron charge- and current densities.
The total time-dependent Hamiltonian $H[\rho^{\bm{\theta}}(t)]$ is a functional of the time-dependent
density operators $\rho^{\bm{\theta}}(t)$ and it is also an explicit function of time through (\ref{ft}). It includes the periodic ring potential (\ref{Vper}), the electron-photon
interaction (\ref{e-gIxIyN}), the electron Coulomb direct and exchange-correlation terms,
the Zeeman spin energy term, the time-dependent modulation of the electron-photon
interaction (\ref{Ht}-\ref{ft}), and the electron and photon
``kinetic terms'' \cite{PhysRevB.110.205301}.
Subsequently, the time-dependent electron and current densities have to be calculated from
the density operator instead of the Fermi distribution as the excitation throws the system out of
equilibrium \cite{PhysRevB.110.205301}.

For the dynamic system we calculate the time-dependent mean photon number
\begin{equation}
    N_\gamma (t) = \frac{1}{(2\pi)^2}\sum_{\sigma}\int^{\pi}_{-\pi} d\bm{\theta}\; \mathrm{Tr} \left\{ \rho^{\bm{\theta}}_{\sigma}(t) a^\dagger_\gamma a_\gamma\right\}
    \label{Ngt}
\end{equation}
and the time-dependent mean orbital magnetization
\begin{equation}
    M_o(t) = \frac{1}{2c{\cal A}}\int_{\cal A} d\bm{r} \left( {\bf r}\times
    \langle {\bf J}({\bf r},t) \rangle \right) \cdot{\bm{e}_z}.
    \label{Mo}
\end{equation}
with ${\cal A} = L^2$. In previous publications we have used the notation $Q_J$ for the orbital
magnetization to emphasize its connection to a moment of the current density.
In addition we evaluate the time-dependent spin magnetization
\begin{equation}
    M_s(t) = -\frac{g\mu^*_\mathrm{B}}{\cal A}\int_{\cal A} d\bm{r} \langle \sigma_z (\bm{r},t)\rangle ,
    \label{Ms}
\end{equation}
with $\sigma_z$ identified as $\zeta = (n_\uparrow - n_\downarrow)/n_\mathrm{e}$, the spin polarization
defined in the spin-DFT formalism. $n_\mathrm{e} = (n_\uparrow + n_\downarrow)$.

The integrand in Eq.\ (\ref{Ngt}) can be used to calculate the photon density in the first Brillouin
zone of the reciprocal space
\begin{equation}
      n_\gamma (\theta_1,\theta_2) = \langle a_\gamma^\dagger a_\gamma \rangle^{\bm{\theta}}
      \label{ngdist}
\end{equation}
for the static or the dynamical system. In the dynamical case the average, or the mean value,
is taken with respect to the density operators as in Eq.\ (\ref{Ngt}) which can be reduced to
diagonal operators in the static system, see Ref.\ \cite{PhysRevB.109.235306}.
{We concentrate our investigation on the three measurable quantities $M_o$,
$M_s$, and $N_\gamma$ for the following reasons: The orbital magnetization in an electron
system in an external magnetic field reflects rotational currents in the system, and these are
sensitive to the circular electric field produced by the TE$_{011}$ quantized photon-cavity mode.
The photon number is not easily accessible in most QEDFT modeling approaches, but in our QED-DFT-TP model it
can be evaluated, and experimentally, it has an important role. The spin magnetization was
included to monitor the consistency of the model, but later became clear that it supplies information about
a spin-phase transition in the system.}

\section{Results}
\label{Results}
For the model calculations, we assume GaAs parameters: the effective electron mass
$m^* = 0.067m_e$, the dielectric constant $\kappa = 12.4$, and the effective
$g$-factor $g^* = -0.44$.
The magnetic flux through a unit cell comes from the commensurability condition
$B{\cal A}= BL^2 =pq\Phi_0$ with the integers $p$ and $q$
\cite{Ferrari90:4598,Silberbauer92:7355,Gudmundsson95:16744},
and the flux quantum $\Phi_0 = hc/e$. For the excitation pulse (\ref{ft})
we use $\hbar\omega_\mathrm{ext} = 3.5$ meV, $\hbar\Gamma = 0.5$ meV, and $V_t/\hbar\omega_c = 0.8$.

\subsection{Two electrons in each ring, $N_\mathrm{e} = 2$}
\begin{figure*}[htb]
    \includegraphics[width=0.48\textwidth]{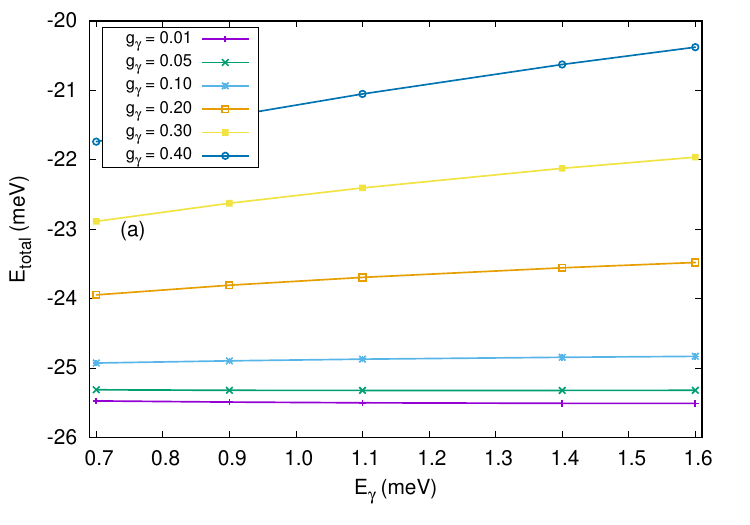}
    \includegraphics[width=0.48\textwidth]{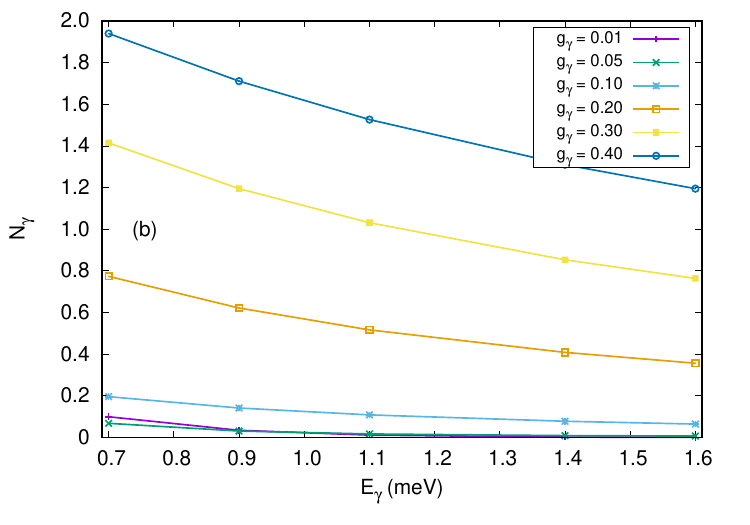}\\
    \includegraphics[width=0.48\textwidth]{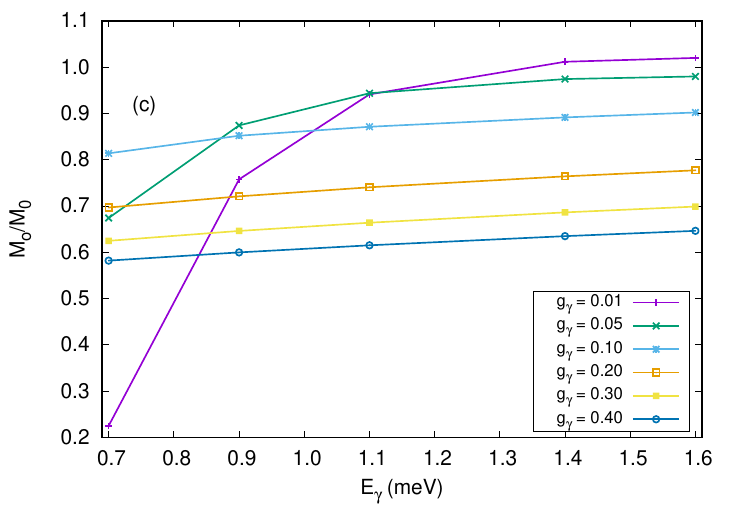}
    \includegraphics[width=0.48\textwidth]{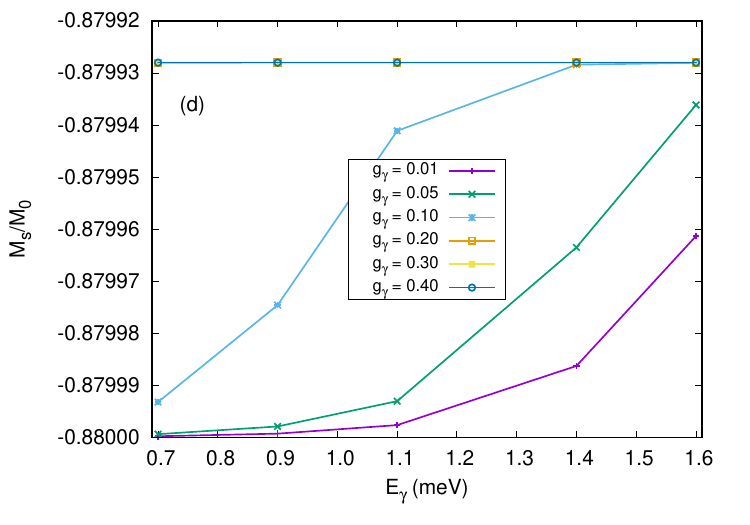}
    \caption{The total energy $E_\mathrm{tot}$ (a), the mean number of photons $N_\gamma$ (b),
             the orbital magnetization $M_o$ (c), and the spin magnetization $M_s$ (d)
             of the 2DEG in a unit cell as function of the photon energy $E_\gamma$
             for different values of the dimensionless electron-photon coupling $g_\gamma$.
             $T = 1.0$ K, $N_\mathrm{e} = 2$, $pq = 2$, and
             $M_0 = \mu^*_\mathrm{B}/L^2$.}
    \label{Yfirlit-pq2-Ne02}
\end{figure*}

Figure \ref{Yfirlit-pq2-Ne02} shows mostly expected results for 2 electrons in each ring. The
total energy $E_{tot}$ increases with increasing electron-photon coupling $g_\gamma$ (a)
and the same can be said about the photon number in each ring $N_\gamma$ (b).
There is a more subtle change in the orbital magnetization $M_o$ (c) at low  photon energy $E_\gamma$
due to the varying occupation of the first photon replica. Importantly, we notice almost no
change in the spin-configuration, seen through the spin magnetization (d), but
a comparison of the results for the array of rings to the results for an array of dots for the same
lattice size, $L = 100$ nm, makes
immediately clear one fundamental difference. In the array of rings the Coulomb exchange interaction
is stronger than for the dot array. It is sufficient to look at Fig.\ 2(d)
in Ref.\ \cite{PhysRevB.109.235306} showing the spin magnetization for $N_\mathrm{e} = 2$ at
the magnetic flux $pq = 2$, corresponding to the magnetic field $B = 0.827$ T. It is very close to 0
for photon energies 1.0 and 2.0 meV and the dimensionless electron-photon coupling $g_\gamma$ in the range
from almost 0 to 0.3. Therefore, the spin configuration of each dot is a singlet state.
For individual dots it has been known for a long time that the singlet state preferred by the system at
low magnetic field changes to a triplet as the magnetic field is increased \cite{Pfannkuche93:2244}.
Note though, that in the calculations for a single dot a positive background charge is usually not assumed,
so the direct Coulomb interaction is stronger than in an array, where a positive background charge
is assumed. Figure \ref{Yfirlit-pq2-Ne02} (in the present paper) shows the total energy
$E_\gamma$, the mean
photon number $N_\gamma$, the orbital $M_o$, and the spin magnetization $M_s$ per unit cell for
$pq = 2$ and $N_\mathrm{e} = 2$.
The spin magnetization in Fig.\ \ref{Yfirlit-pq2-Ne02}(d) shows that the two electrons in each ring
are in a spin triplet state, even for very high electron-photon coupling.
We find the same happening for the lower magnetic flux $pq = 1$ (not shown here) for the array
of rings, the electrons remain in the triplet spin state.

For the dot system, i.e.\ an array
of quantum dots, all our previous calculations using the QED-DFT-TP \cite{PhysRevB.109.235306},
or the QEDFT \cite{PhysRevB.106.115308} show clearly that within the parameter range used here,
the increased electron-photon interaction weakens the Coulomb exchange interactions considerably.
{Like for the dots, the increased electron photon coupling tends to polarize (smear) the
charge density radially outward from the center of the rings enhancing their orbital magnetization}.

Below, we check better the static and dynamic properties of the ring array with $N_\mathrm{e} = 2$
and $pq = 2$ flux quanta flowing through the unit cell before analyzing the situation for $N_\mathrm{e} = 3$.

\begin{figure}[htb]
    \includegraphics[width=0.48\textwidth]{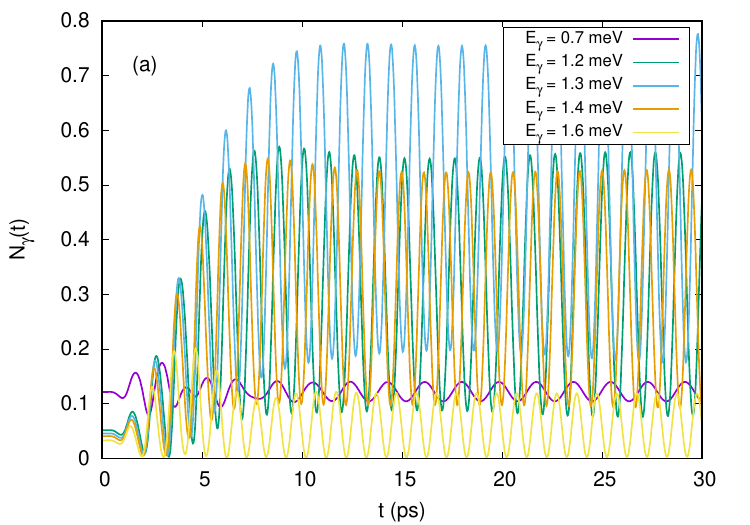}
    \includegraphics[width=0.48\textwidth]{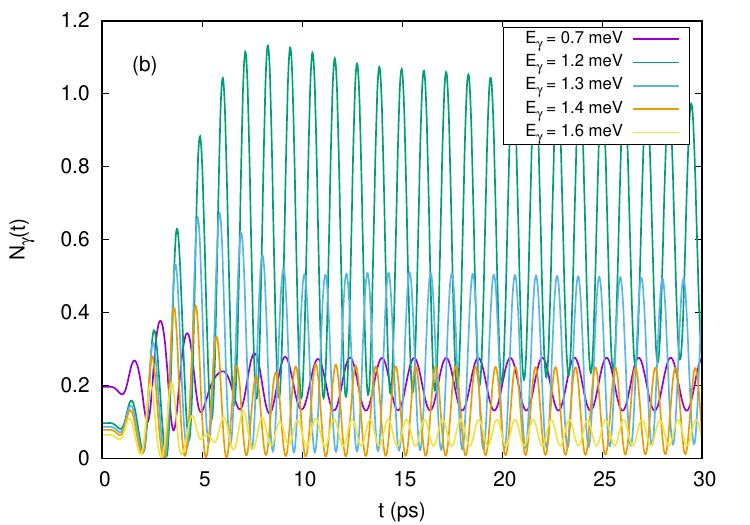}
    \caption{The dynamic mean photon number $N_\gamma (t)$ per unit cell of the
             2DEG as a function of time for the first 30 ps for different values
             of the photon energy $E_\gamma$ and $g_\gamma = 0.08$ (a), and
             the electron-photon coupling $g_\gamma = 0.10$ (b).
             $N_\mathrm{e} = 2$, $pq = 2$.}
    \label{Ng-pq2-Ne02}
\end{figure}

After the same dynamical excitation (\ref{Ht}) as is used in
Ref.\ \cite{PhysRevB.110.205301}
the mean photon number oscillates as is displayed in Fig.\ \ref{Ng-pq2-Ne02} for
$g_\gamma = 0.08$ in (a), and $0.10$ in (b).
The oscillations are similar as for the quantum dot array, except for the observation
that the excitation of the ring array looks to be more effective. That is not surprising as
the active states of the ring system, in a polar coordinate system not used here, have mainly
contributions from the eigenstates of the angular momentum operator for
finite angular momentum, $M\neq 0$. In a quantum dot the ground state has a higher contribution
from a state with zero angular momentum, the $M=0$ state, in polar coordinates, that is less
susceptible to a circular symmetric excitation. In polar coordinates $M$ is the
angular momentum quantum number.

\begin{figure}[htb]
    \includegraphics[width=0.48\textwidth]{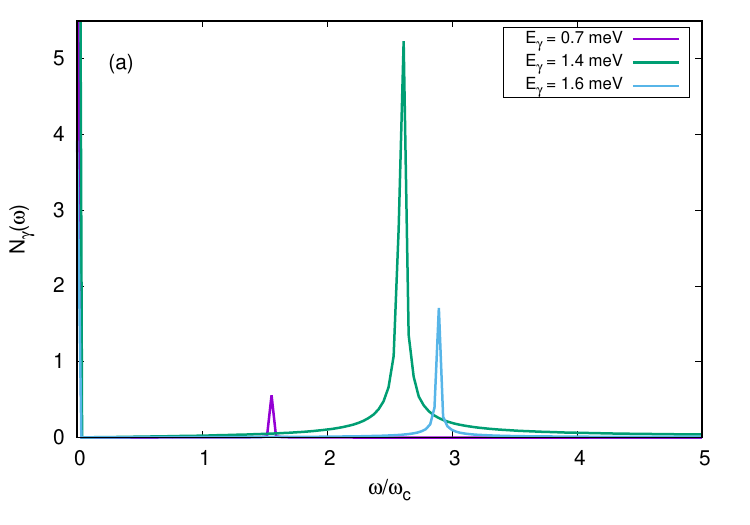}
    \includegraphics[width=0.48\textwidth]{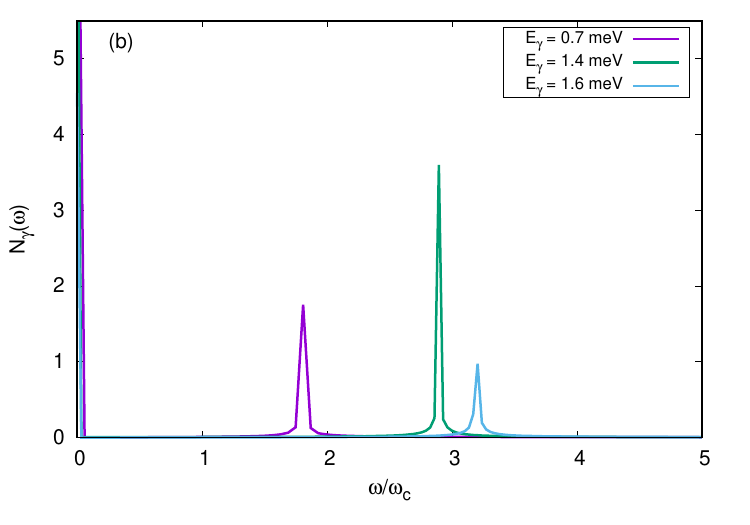}
    \caption{The Fourier power spectra of the mean photon number per unit cell
             $N_\gamma (\omega )$ for different values of the photon energy
             $E_\gamma$ and the electron-photon coupling $g_\gamma = 0.08$ (a),
             and $g_\gamma = 0.10$ (b).
             $N_\mathrm{e} = 2$, $pq = 2$, and $\hbar\omega_c = 1.429$ meV.}
    \label{P-pq2-Ne02}
\end{figure}

The Fourier power spectra for the two-electron rings are displayed in Fig.\ \ref{P-pq2-Ne02}.
The only difference between the excitation spectra is that in Fig.\ \ref{P-pq2-Ne02}(a)
the coupling $g_\gamma = 0.08$, but it is $0.10$ in (b). We notice the peaks are not at the
bare photon energies, but are shifted proportionally to the electron-photon coupling
$g_\gamma$. 

This interaction shift is exactly the reason for which excitation peak
is best excited by the frequency spectrum the excitation forces on the system
\cite{PhysRevB.110.205301}. The spectrum of the excitation is the same for
Fig.\ \ref{P-pq2-Ne02}(a) and (b). 
{
This interaction shift together with the photon energy pushes the excitation
peaks in to, or out of, the maximally effective frequency range of the 
external excitation pulse ($2 < \omega/\omega_c < 3)$}. 

\subsection{Three electrons in each ring, $N_\mathrm{e} = 3$}
\begin{figure}[htb]
    \includegraphics[width=0.48\textwidth,bb= 20 60 370 265]{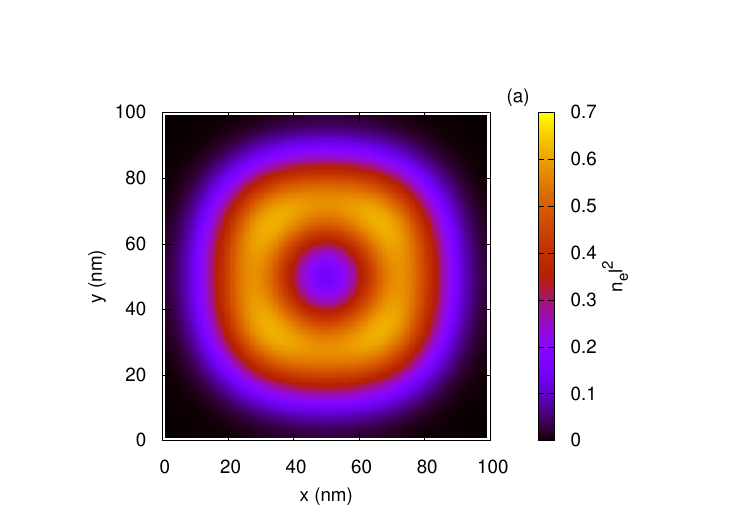}
    \includegraphics[width=0.48\textwidth,bb= 20 60 370 265]{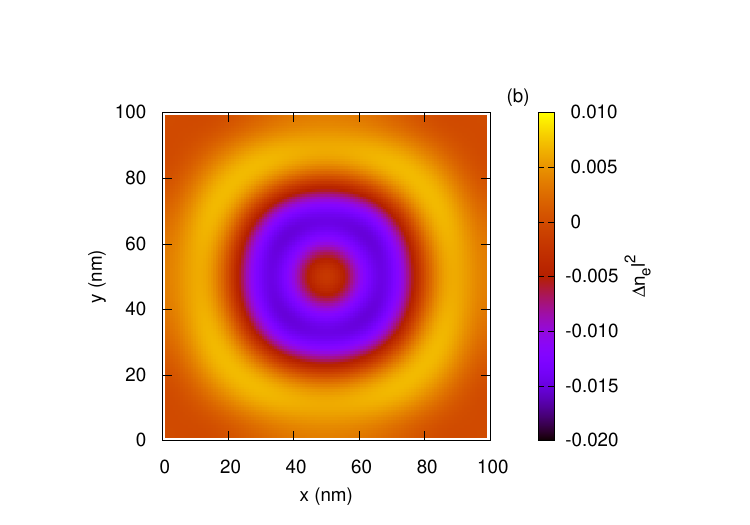}
    \includegraphics[width=0.48\textwidth,bb= 20 60 370 265]{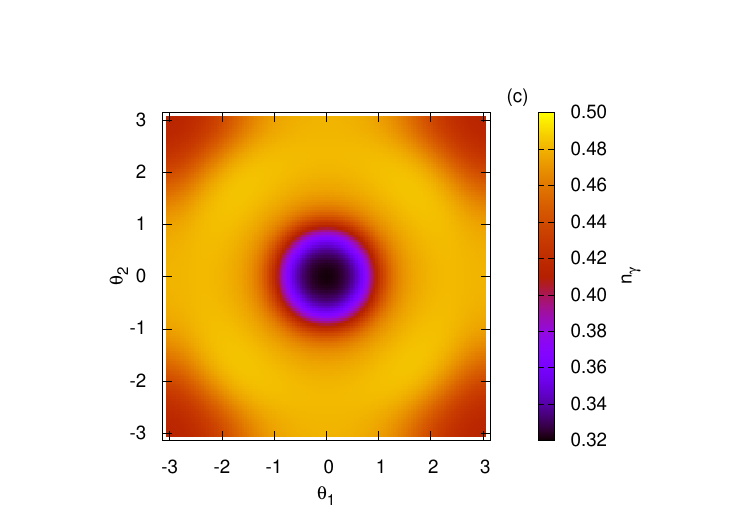}
    \vspace*{0.8cm}
    \caption{The electron density $n_\mathrm{e}l^2$ (a), the change in the electron density
             $\Delta n_\mathrm{e}l^2 =
             [n_\mathrm{e}(g_\gamma = 0.10) -  n_\mathrm{e}(g_\gamma = 0.01)]l^2$ (b),
             and the mean photon distribution $n_\gamma (\theta_1,\theta_2)$
             in the reciprocal unit cell (c), see Eq.\ (\ref{ngdist}).
             $g_\gamma = 0.10$, $N_\mathrm{e} = 3$, $pq = 2$, $E_\gamma = 0.70$ meV,
             $l = 28.21$ nm.}
    \label{d-pq2-Ne03}
\end{figure}

Here, we will analyze the array of quantum rings with $N_\mathrm{e} = 3$ electrons in each ring
and $pq = 2$ flux quanta through the unit cell. First, in Fig.\ \ref{d-pq2-Ne03} we observe
through the electron density (a) and the photon induced change of it (b), that we are dealing with
a ring formed electron system in each unit cell of the superlattice.
In addition, we see the distribution of the photons in the first Brillouin zone of the
reciprocal space (c).
The same is true
for two electrons at $pq = 2$. The ring structure is not as clear for 2 or 3 electrons
at the lower magnetic flux $pq = 1$. There, the hole in the center of the charge distribution
is not quite as deep as for $pq = 2$.
That can be explained by the magnetic length $l$, that
is $28.21$ nm at $pq = 2$, but $39.89$ nm at $pq = 1$. We notice in Fig.\ \ref{d-pq2-Ne03}(b)
that the electron density is polarized radially outwards as the electron-photon coupling
is increased. This is in accordance with our results for dot arrays, both in the QEDFT
\cite{PhysRevB.106.115308} and the QED-DFT-TP \cite{PhysRevB.109.235306} formalism.
Of course, the symmetry of the square array can distort the charge polarization, we see a slight
indication of that here.

{The electron density seen in Fig.\ \ref{d-pq2-Ne03} suggests that the inter-ring
Coulomb exchange is enhanced for an array of rings due to the ``closer spacing'' of
the charge in ring arrays than dot arrays for the case of few electrons in each ring or dot. 
Interestingly, increased electron-photon coupling
slightly reduces the effective distance between the charge in the rings, if it is measured as the distance
between the maxima in the charge distribution of the rings}.

\begin{figure*}[htb]
    \includegraphics[width=0.48\textwidth]{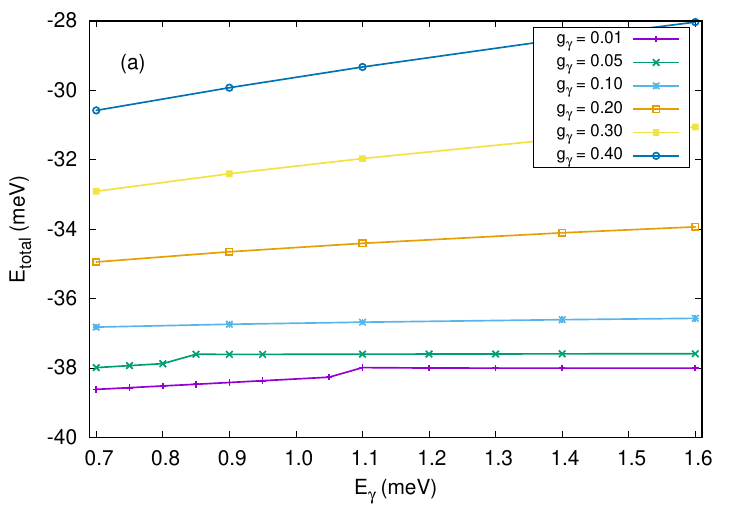}
    \includegraphics[width=0.48\textwidth]{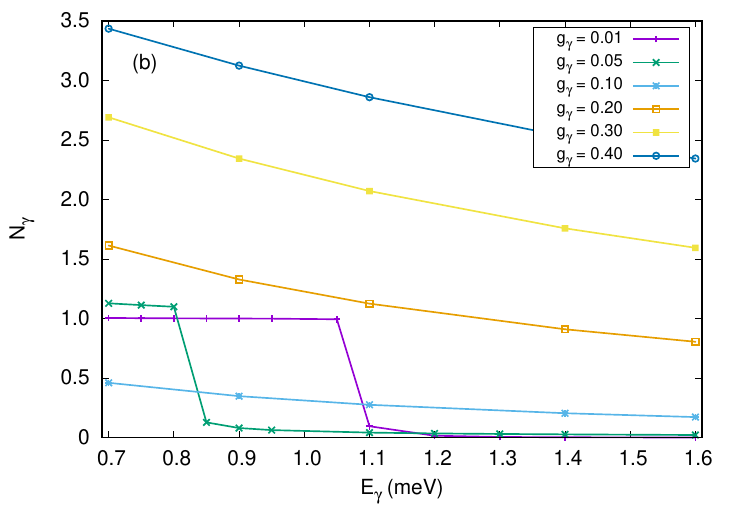}\\
    \includegraphics[width=0.48\textwidth]{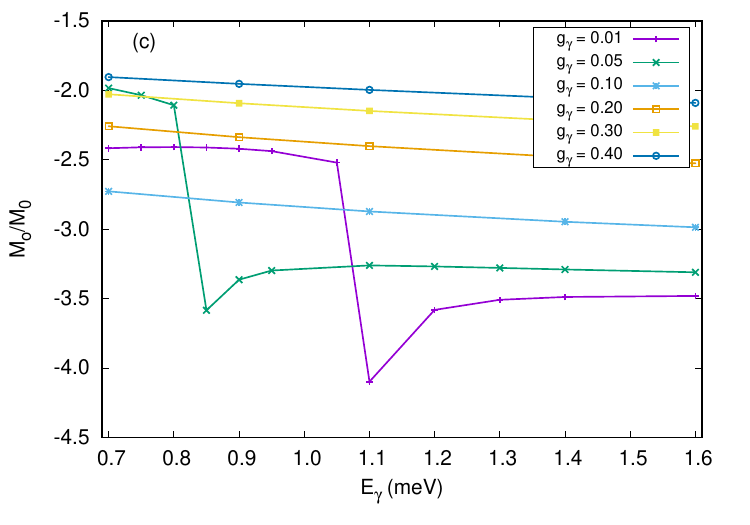}
    \includegraphics[width=0.48\textwidth]{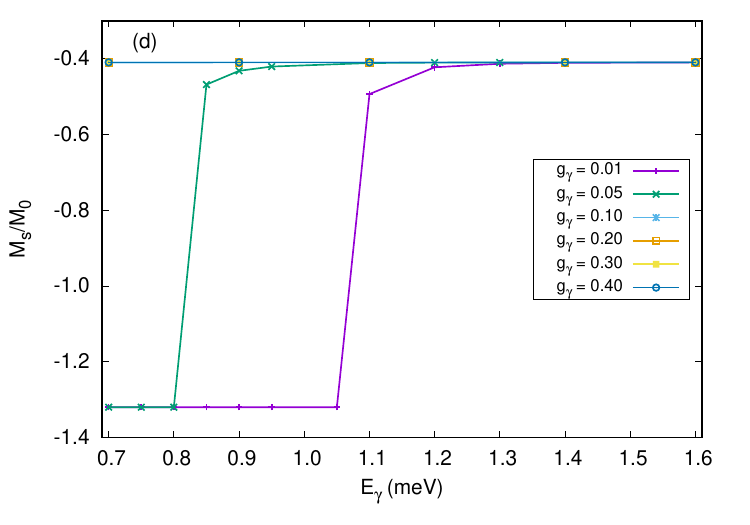}
    \caption{The total energy $E_\mathrm{tot}$ (a), the mean number of photons $N_\gamma$ (b),
             the orbital magnetization $M_o$ (c), and the spin magnetization $M_s$ (d)
             of the 2DEG in a unit cell as function of the photon energy $E_\gamma$
             for different values of the dimensionless electron-photon coupling $g_\gamma$.
             $T = 1.0$ K, $N_\mathrm{e} = 3$, $pq = 2$,
             $M_0 = \mu^*_\mathrm{B}/L^2$.}
    \label{Yfirlit-pq2-Ne03}
\end{figure*}

Figure \ref{Yfirlit-pq2-Ne03} shows the total energy $E_\mathrm{tot}$, the mean
photon number $N_\gamma$, the orbital $M_o$, and the spin magnetization $M_s$ per unit cell for
$pq = 2$ and $N_\mathrm{e} = 3$. Figure \ref{Yfirlit-pq2-Ne03} should be compared
to Fig.\ \ref{Yfirlit-pq2-Ne02}
for two electrons in each ring. We observe a clear phase change from the fully spin polarized state
to a phase of single unpaired spins in each ring, when either the photon energy $E_\gamma$,
or the electron-photon coupling $g_\gamma$ is increased for low photon energy.
Signs of this phase change in the static ring array are seen in all the variables displayed in
Fig.\ \ref{Yfirlit-pq2-Ne03}, though they are clearest in subfigure (d) for the spin magnetization
$M_s$. {For a high photon energy the orbital magnetization increases with increasing
electron-photon interaction as it polarizes the electron charge radially outwards from the center of
the rings. The situation is more complex for lower photon energy as there exact configuration of
the electron states matters, and it changes with the spin structure. The photon number generally tends
to decrease with increasing photon energy due to the location of the respective photon replica states
higher in the energy spectrum.}

We have done a calculation for the lower magnetic flux of $pq = 1$ for 3 electrons.
There, for all values of the photon energy, we find the system to be maximally spin
polarized for low electron-photon coupling, but for all other values of higher
coupling, only a single unpaired spin is found in each ring.

\begin{figure*}[htb]
    \includegraphics[width=0.48\textwidth]{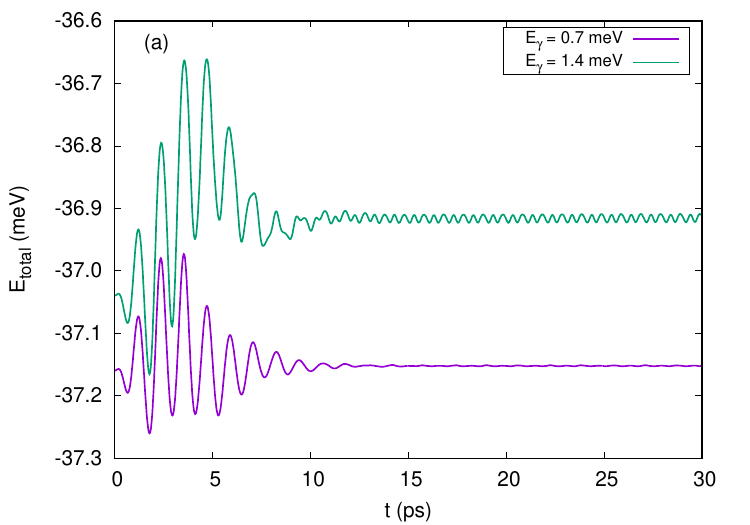}
    \includegraphics[width=0.48\textwidth]{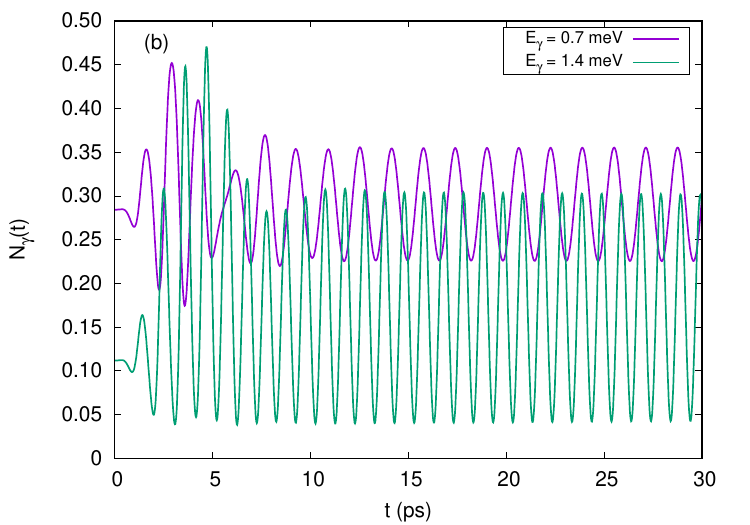}\\
    \includegraphics[width=0.48\textwidth]{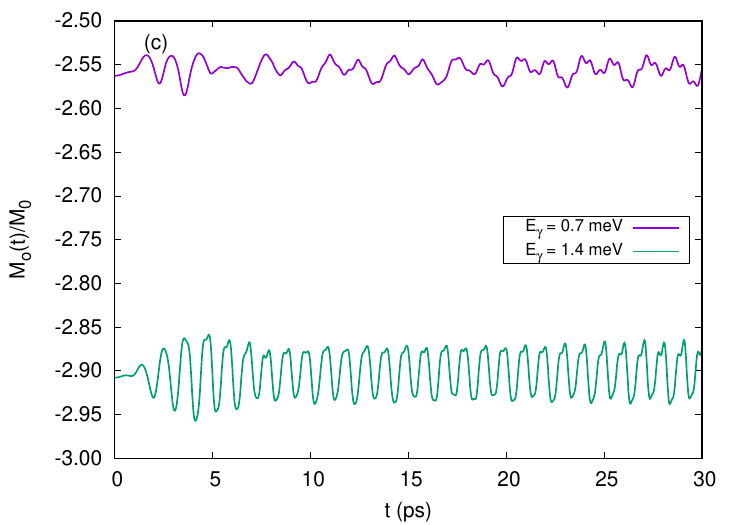}
    \includegraphics[width=0.48\textwidth]{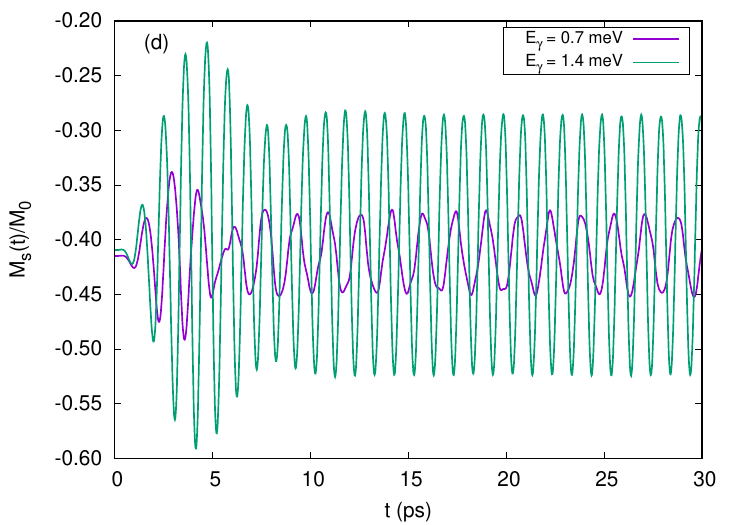}
    \caption{The total energy $E_\mathrm{tot}$ (a), the mean number of photons $N_\gamma$ (b),
             the orbital magnetization $M_o$ (c), and the spin magnetization $M_s$ (d)
             of the 2DEG in a unit cell as function of time for the first 30 ps
             for different values of the photon energy $E_\gamma$.
             $g_\gamma = 0.08$, $N_\mathrm{e} = 3$, $pq = 2$, and
             $M_0 = \mu^*_\mathrm{B}/L^2$.}
    \label{Yt-pq2-Ne03-gg008}
\end{figure*}

These signs of a spin-phase transition in the static system open the question if any signs of it
can be observed in the dynamic evolution of the system after an excitation.
Indeed, signs can be found. The excitation of the array of quantum rings should not
promote a change in the mean spin configuration. It can change the mean total orbital momentum, or
the orbital magnetization in each ring with the induced circular electric field, but the spin
should be conserved, but the mean spin in each ring might fluctuate around its original value.
This is indeed what happens as can be seen in Fig.\ \ref{Yt-pq2-Ne03-gg008} for $g_\gamma = 0.08$.
Additionally, we see a slight trend to a period doubling in the orbital magnetization, reflecting
an increase of the higher order transitions being activated.

\begin{figure*}[htb]
    \includegraphics[width=0.48\textwidth]{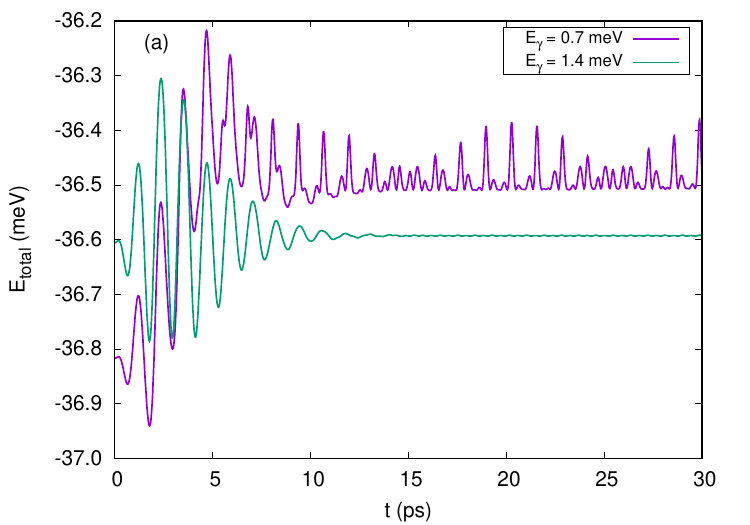}
    \includegraphics[width=0.48\textwidth]{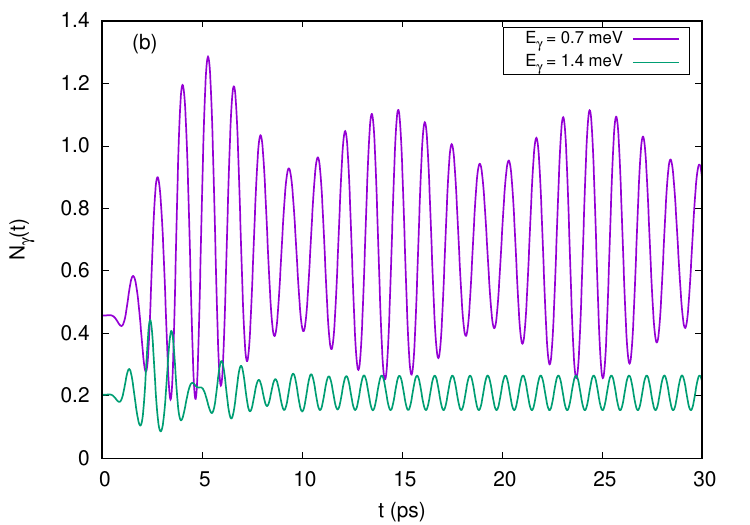}\\
    \includegraphics[width=0.48\textwidth]{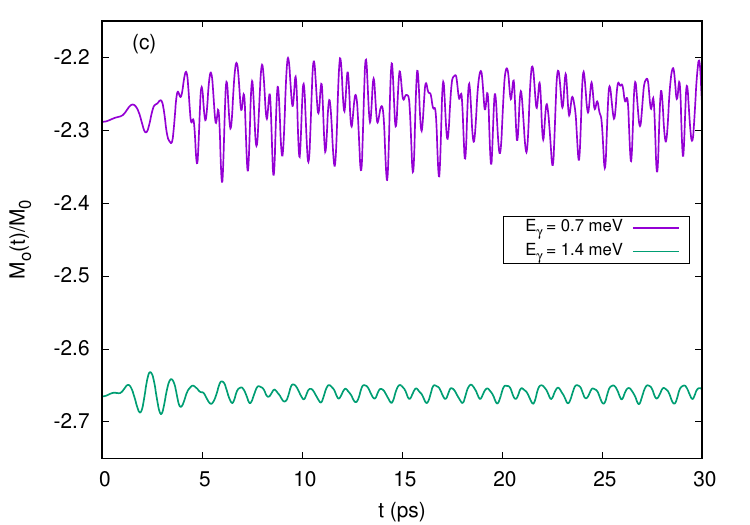}
    \includegraphics[width=0.48\textwidth]{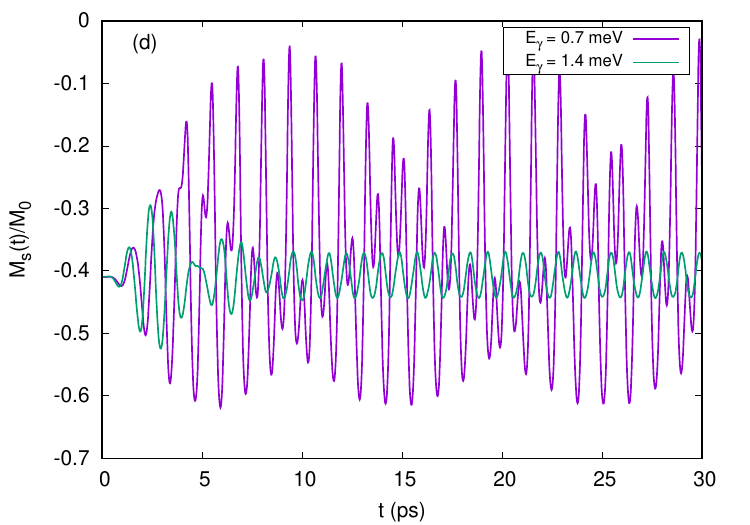}
    \caption{The total energy $E_\mathrm{tot}$ (a), the mean number of photons $N_\gamma$ (b),
             the orbital magnetization $M_o$ (c), and the spin magnetization $M_s$ (d)
             of the 2DEG in a unit cell as function of time for the first 30 ps
             for different values of the photon energy $E_\gamma$.
             $g_\gamma = 0.10$, $N_\mathrm{e} = 3$, $pq = 2$, and
             $M_0 = \mu^*_\mathrm{B}/L^2$.}
    \label{Yt-pq2-Ne03-gg010}
\end{figure*}

The same behavior with more extremes can be seen for the case of the stronger electron-photon
coupling $g_\gamma = 0.10$ displayed in Fig.\ \ref{Yt-pq2-Ne03-gg010}.
{Here, we encounter beating patterns that will be addressed shortly below.
Most importantly, the spin fluctuation is stronger than in Fig.\ \ref{Yt-pq2-Ne03-gg008}
and clearly the energy pumped into the system is now larger for the lower photon energy,
in contrast to what is seen in Fig.\ \ref{Yt-pq2-Ne03-gg008}. This again can be explained by
the location of the excitation peaks with respect to the spectrum of the initial excitation pulse.}

\begin{figure*}[htb]
    \includegraphics[width=0.48\textwidth]{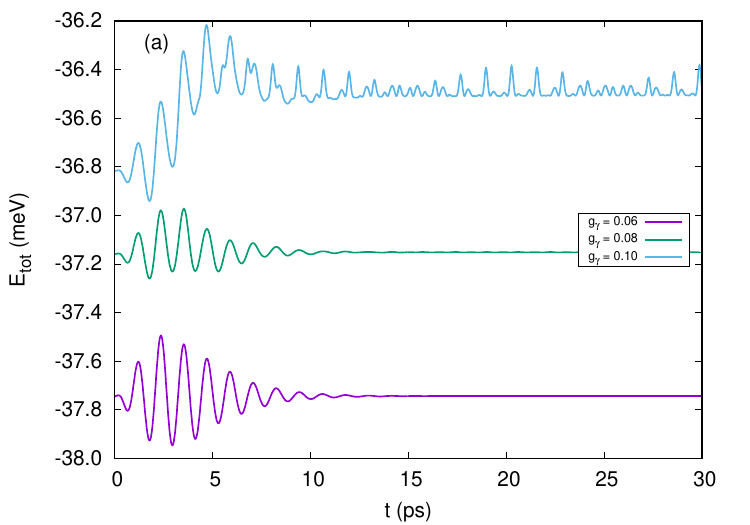}
    \includegraphics[width=0.48\textwidth]{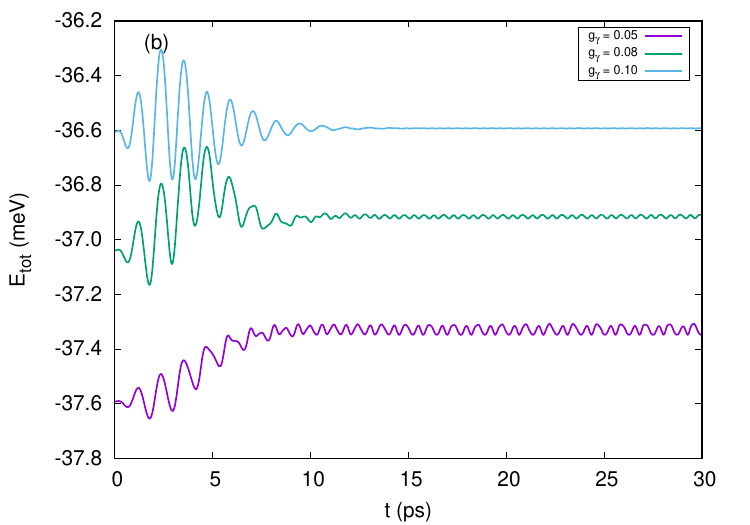}\\
    \includegraphics[width=0.48\textwidth]{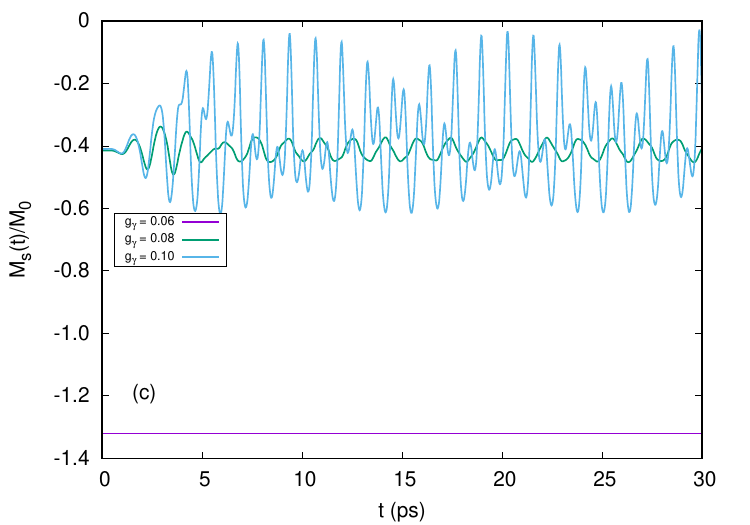}
    \includegraphics[width=0.48\textwidth]{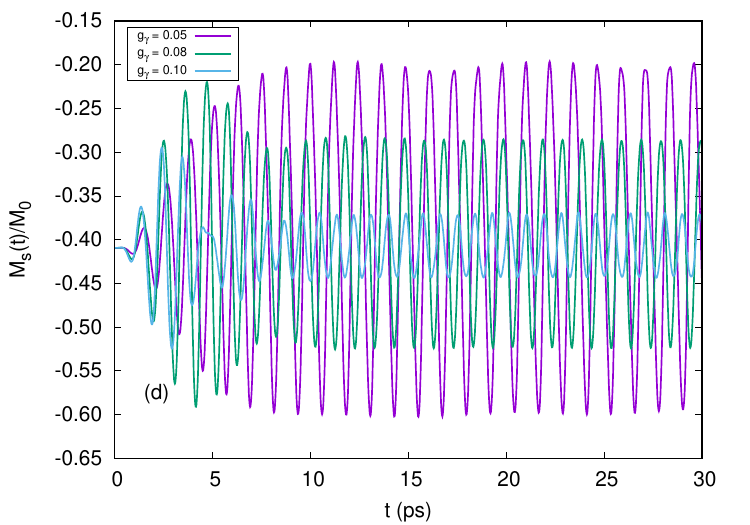}
    \caption{The total energy $E_\mathrm{tot}$ for the photon energy $E_\gamma = 0.70$ meV (a),
             $E_\gamma = 1.40$ meV (b), and
             the spin magnetization $M_s$ for $E_\gamma = 0.70$ meV (c), $E_\gamma = 1.40$ (d)
             of the 2DEG in a unit cell as function of time for the first 30 ps
             for different values of the electron-photon coupling $g_\gamma$.
             $N_\mathrm{e} = 3$, $pq = 2$, and $M_0 = \mu^*_\mathrm{B}/L^2$.}
    \label{Ms-pq2-Ne03-gg}
\end{figure*}

In Fig.\ \ref{Ms-pq2-Ne03-gg} this is compared for the total energy and the spin-magnetization for
two different values of the photon energy, and three values of the electron-photon
coupling.
It is particularly interesting to see the difference in the spin-magnetization in
subfigure (c). For the low electron-photon coupling of $g_\gamma = 0.06$ no fluctuations
are seen on the scale used to capture the fluctuations for the other values of the coupling.

We stress that the fluctuations in the spin magnetization only show excitation peaks
in the same energy locations as the mean photon number and the dynamic orbital magnetization.
This reflects the fact that the chosen excitation can not exert any torque on the spins.

{The beating patterns visible in Figures \ref{Yt-pq2-Ne03-gg010} and \ref{Ms-pq2-Ne03-gg}, and
to some extent also in Fig.\ \ref{Yt-pq2-Ne03-gg008}, especially for the spin magnetization $M_s(t)$,
are caused by interference of the fundamental peak of $M_s(\omega)$ and the slightly shifted second harmonics
thereof. The shift of both peaks is due to the electron photon interaction, 
but is also influenced by the band structure of the electron-photon system.}

We specially note that we do not observe any formation of a spin density wave in the
angular direction within one quantum ring, neither in the static nor in the dynamic system.
The total energy in Fig.\ \ref{Yt-pq2-Ne03-gg008}(a), \ref{Yt-pq2-Ne03-gg010}(a),
and \ref{Ms-pq2-Ne03-gg}(a) and (b) is not totally constant after the excitation pulse ends
at $t = 16$ ps. This has to be understood in relation with the Heisenberg uncertainty laws.
The average of the total energy oscillates around a constant value. The number of electrons
within a unit cell of the periodic system and the trace of the density matrices in each point
$\bm{\theta}$ retain their correct values with 11 to 12 significant digits.

\section{Conclusions}
\label{Conclusions}
We have used a QED-DFT-TP formalism to describe the Coulomb interaction and a configuration 
interaction (CI) approach for
both the para- and the diamagnetic electron-photon interactions for a 2DEG in a large
superlattice of quantum rings embedded in a cylindrical FIR-photon cavity in an external
magnetic field.

We find that in the ring array the Coulomb exchange interaction is stronger than in the
corresponding array of quantum dots. For two electrons in each dot we observe a spin triplet
state in each ring, even at low magnetic field.

For three electrons and two magnetic flux quanta $pq = 2$ in each ring cell we find a fully spin
polarized state for the static system for low electron-photon coupling and small photon energy.
As the electron-photon coupling, or the photon energy, are increased a phase transition to a single
unpaired spin in each cell is observed in the superlattice. For a single magnetic flux
quantum $pq = 1$ through the unit cell we observe the same kind of a spin-phase transition
with respect to the strength of the electron-photon coupling, whereas
a change in the photon energy in the range we use does not lead to a phase transition.

In the system with $pq = 2$ and two electrons in each ring a dynamical excitation
of the system does not change its spin configuration, but tiny oscillations in the
mean spin magnetization, most often just close to the numerical noise level, can be
observed. The same could be said for all our former calculations for the excitation
of arrays of quantum dots within the same formalism. The excitation does not
affect the spin degree of freedom as analytical evaluations show, but only the
population of different states.
For the 3-electron rings in the $pq = 2$ magnetic flux, we observe spin fluctuations
in the system close to a phase transition, even though the mean spin value is conserved.

We observe thus, directly a spin-phase transition in the static system, but in the
dynamical system we see the fluctuations that are known to be present in systems
close to phase transitions. Both these phenomena support the idea that the superlattice,
or array, of rings shows in this case a genuine spin-phase transition, and that the inter-ring
exchange interaction is at work. Furthermore, the spin-phase transition can be suppressed by
a strong electron-photon interaction. This suppression of the Coulomb exchange has been
found previously in different models of corresponding quantum dots arrays.

{
The fact that we observe a spin-phase transition for the ring array with 3 electrons in
each ring, but not when each ring is occupied with 2 electrons within the parameters space
covered in our calculations points out, not surprisingly, that the spin structure of a 3 electron system 
is more fragile}.

{
We are aware that even though we manage to find temporal fluctuations of the spin configuration
in our model, a strictly periodic hybrid DFT-CI model is not adequate to state anything about
possible spatial fluctuations and diverging length scales that should be expected at the phase-transition. {\it We do thus not claim to observe a genuine thermodynamic (quantum) phase transition with corresponding universal features or diverging behaviors}}.

Finally, let us point out that the arrays of quantum dots or rings placed in photon cavities are in
fact large and extended hybrid quantum systems, in contrast to smaller {\it qubit-like}
devices (e.g.\ quantum dots coupled to microwave cavities \cite{PhysRevX.7.011030,PhysRevX.12.031004}
or color centers coupled by phononic modes \cite{PhysRevLett.132.263602,PhysRevX.9.031045}).
The latter are extensively used in quantum technologies, whereas the former
provide important platforms merging collective effects and strong field-matter interaction.
Note also that for large hybrid systems the theoretical tools and modeling are quite involved,
as already illustrated in our present and previous studies based on QED-DFT and QED-DFT-TP methods.

\begin{acknowledgments}
This work was financially supported by the Research Fund
of the University of Iceland grant No.\ 92199, and the Icelandic Infrastructure Fund
for ``Icelandic Research e-Infrastructure (IREI)''.
The computations were performed on resources
provided by the Icelandic High Performance Computing
Center at the University of Iceland.
V.\ Mughnetsyan and V.\ Gudmundsson acknowledge support
by the Higher Education and Science Committee of Armenia (grant No.\ 24LCG-1C004).
V.\ Gudmundsson acknowledges support for his visit to the National Taiwan University from the National Science and Technology Council, Taiwan under Grants No.\ NSTC 113-2811-M-002-001
and No.\ NSTC 112-2119-M-002-014.
H.-S.G.\ acknowledges support from the National Science and Technology Council, Taiwan under Grants No.\ NSTC 113-2112-M-002-022-MY3, No.\ NSTC 113-2119-M-002 -021, No.\ NSTC 112-2119-M-002-014, No.\ NSTC 111-2119-M-002-007, and No.\ NSTC 111-2627-M-002-001, from the US Air Force Office of Scientific Research under Award Number FA2386-23-1-4052 and from the National Taiwan University under Grants No.\ NTU-CC-113L891604 and No.\ NTU-CC-114L8950. H.-S.G.\ is also grateful for the support from the ``Center for Advanced Computing and Imaging in Biomedicine (NTU-113L900702)'' through The Featured Areas Research Center Program within the framework of the Higher Education Sprout Project by the Ministry of Education (MOE), Taiwan, and the support from the Physics Division, National Center for Theoretical Sciences, Taiwan.
J.-D.\ Chai acknowledges support from the National Science and Technology Council, Taiwan under
Grant No.\ NSTC 113-2112-M-002-032. J.-D.\ Chai is also grateful for the support from the Physics Division,
National Center for Theoretical Sciences, Taiwan.
C.-S.\ Tang acknowledges funding support by the National United
University through Contract No.\ 113-NUUPRJ-01.
V.\ Moldoveanu acknowledges financial
support from the Core Program of the National Institute of Materials Physics, granted by the Romanian Ministry
of Research, Innovation and Digitalization under the Project PC2-PN23080202.

\end{acknowledgments}

%
\appendix
\section{Technical details in the methodology}
\label{Tech-details}

For the calculations we use all the same parameters as is detailed in Appendix B in
Ref.\ \cite{PhysRevB.110.205301}, except that in Figures
\ref{Yfirlit-pq2-Ne02}, \ref{d-pq2-Ne03}, and \ref{Yfirlit-pq2-Ne03}, we use the
16 lowest eigenstates of the photon number operator
and a $32\times 32$ nonequispaced grid in the first Brillouin zone of the reciprocal space, $\bm{\theta}$,
built on a repeated 4-point Gaussian quadrature. This is done to guarantee the accuracy of the results for
the static system for the electron-photon coupling $g_\gamma > 0.10$.

The ground state, or static, calculations are done in a linear functional basis constructed as a
tensor product (TP) of electron and photon states
$|\bm{\alpha\theta}\sigma n\rangle = |\bm{\alpha\theta}\sigma\rangle\otimes|n\rangle$.
The electron states are states proposed by Ferrari \cite{Ferrari90:4598}, and used by
Gudmundsson \cite{Gudmundsson95:16744} and Silberbauer \cite{Silberbauer92:7355}.
The photon states are the eigenstates of the photon number operator
$N_\gamma = a^\dagger_\gamma a_\gamma$ with eigenvalue $n$. $\sigma$ is the spin label
$\{\uparrow\downarrow\}$, and $\bm{\alpha}$ is a composite quantum number for the Landau subband number.

The self-consistent diagonalization of the static total Hamiltonian leads to the states
$|\bm{\alpha\theta}\sigma n)$, but the Liouville-von Neumann Eq.\ (\ref{L-vN}) is solved in
the $\{|\bm{\alpha\theta}\sigma n\rangle\}$-basis as most matrix elements are known analytically there
\cite{Gudmundsson95:16744,PhysRevB.110.205301}.
%

%

%

\end{document}